\documentclass[twocolumn,superscriptaddress]{revtex4}
\usepackage{graphicx}
\usepackage{epsfig}
\usepackage{amsmath,amssymb}
\usepackage{amsfonts,amsbsy}
\usepackage{hyperref}
\usepackage{bm}
\usepackage{color}
\usepackage{enumerate}
\usepackage{verbatim}
\usepackage{lipsum}

\def\be{\begin{equation}}
\def\ee{\end{equation}}
\def\bea{\begin{eqnarray}}
\def\eea{\end{eqnarray}}

\newcommand{\bo}[1]{\boldsymbol{#1}}

\begin{document}
\title{Large-rapidity ridge correlations from Color Glass Condensate}

\author{Donghai Zhang}
\affiliation{Key Laboratory of Quark and Lepton Physics (MOE) and Institute of Particle Physics, Central China Normal University, Wuhan 430079, China}
\affiliation{School of Physics and Astronomy, China West Normal Univeristy, Nanchong 637002, China}
\author{Yeyin Zhao}
\affiliation{Key Laboratory of Quark and Lepton Physics (MOE) and Institute of Particle Physics, Central China Normal University, Wuhan 430079, China}
\author{Mingmei Xu}
\email{xumm@ccnu.edu.cn}
\affiliation{Key Laboratory of Quark and Lepton Physics (MOE) and Institute of Particle Physics, Central China Normal University, Wuhan 430079, China}
\author{Yuanfang Wu}
\email{wuyf@ccnu.edu.cn}
\affiliation{Key Laboratory of Quark and Lepton Physics (MOE) and Institute of Particle Physics, Central China Normal University, Wuhan 430079, China}
\date{\today}
\begin{abstract}
Within the Color Glass Condensate (CGC) effective field theory, considering the violation of boost invariance of the rapidity distribution, we correct the normalization scheme of the longitudinal rapidity ridge correlations. After this correction, the large-rapidity ridge correlation rebounds after bottoming, consistent with the observed data from the CMS detector. It is also found that the correlation rebound appears around the sum of the saturation momentum of the projectile and target, and moves to larger rapidities at higher collision energies. These features directly result from the saturation and the quantum evolution of gluons within the framework of the CGC. 
\end{abstract}

\maketitle
\section{Introduction}
Two-particle correlations are powerful observables in exploring the underlying mechanisms of particle production. The $\Delta\eta$-$\Delta\phi$ correlation functions for three kinds of collisions, i.e. proton-proton~\cite{CMS-2010-pp,CMS-2011-pp-7,ATLAS-2015-pp-13,CMS-2016-pp-13,ALICE-2017-pp-7,CMS-2017-pp-13}, proton-nucleus~\cite{CMS-2013-pPb-5.02,ALICE-2013-pPb-5.02,ATLAS-2013-pPb-5.02,ATLAS-2013-pPb-5.02-2,CMS-2013-pPb-PbPb,ALICE-2016-pPb-5.02} and nucleus-nucleus collisions~\cite{CMS-2011-pp-PbPb,CMS-2011-PbPb-2.76,CMS-2012-PbPb-2.76,PHENIX-2008,STAR-2009,PHOBOS-2010}, show similar structures. There is an enhancement on the near side (relative azimuthal angle $\Delta\phi\approx0$) that extends over a wide range in relative pseudorapidity ($|\Delta\eta|\approx4$). In particular, a rebound at $|\Delta\eta|\approx4$ after a plateau within $2<|\Delta\eta|<3.6$ is observed~\cite{CMS-2011-pp-7}. Such a long-range near-side correlation is known as the ``ridge". This ridge like feature has drawn a lot of attentions in both experimental and theoretical sides. 

At present, there are two mainstream mechanisms to explain the ridge in small systems, namely the glasma correlation in the initial state~\cite{Dumitru-2008,Dusling-2010,Dumitru-2011,Dusling-2012,Dusling-2013-1,Dusling-2013-2,Dusling-2013-3,Mace-2019} described by the CGC effective field theory and the final state evolution~\cite{Werner-2011,Bozek-2013,Bzdak-2013,Qin-2014,Werner-2014} described by hydrodynamics.

On one hand, hydrodynamics systematically described azimuthal anisotropies in small systems including p-Au, d-Au and He-Au collisions at RHIC and p-Pb collisions at LHC~\cite{Werner-2011,Bozek-2013,Bzdak-2013,Qin-2014,Werner-2014}. It fails to reproduce multiple-particle cumulant $c_2\{4\}$~\cite{hydro-c24} and the elliptic flow of heavy flavor particles~\cite{hydro-flavor}. On the other hand, CGC successfully explained the mass ordering of $v_2$~\cite{CGC-mass}, multiple-particle cumulant $c_2\{4\}$~\cite{CGC-c24}, the elliptic flow of heavy flavor particles~\cite{CGC-flavor} and $\gamma$A process~\cite{CGC-gammaA}, except the ordering of Fourier harmonics $v_n$ with respect to system sizes of p-Au, d-Au and He-Au~\cite{CGC-fail}. Recent research~\cite{latest-1,latest-2} shows that the azimuthal anisotropy measured in small systems may be the combined contributions of CGC and hydrodynamics. 


The CGC effective field theory provides a consistent description for collisions of both hadrons and nuclei~\cite{CGC-review}. The gluon density inside a colliding hadron or nucleus grows with the collision energy. Gluon saturation occurs at high enough energy or small enough Bjorken $x$. The colliding hadrons of high gluon density are two sheets of color glass condensate which can be described by classical color electric and color magnetic fields. When two sheets of color glass pass through each other, the high intensity color fields interact and evolve. After the collision, strong longitudinal color electric and color magnetic fields are formed in the region between the colliding hadrons, which is called glasma~\cite{glasma}. The approximate longitudinal boost invariant glasma fields produce particles with long range rapidity correlations. 

The glasma field is an equivalent description of dense gluons in the initial state. The quantum evolution of projectiles~\cite{Dusling-2010} starts from the radiations of valence quarks and produces gluons. Successive gluon splittings produce more gluons. Gluon recombination stops the increase of the gluon density. The gluon density saturates finally. 

For a right moving projectile, the Bjorken $x$ of a gluon reads
\begin{equation}
x=\frac{p_\perp}{\sqrt{s}}{\rm e}^{y},
\end{equation}
with $y$ representing rapidity, $p_\perp$ transverse momentum and $\sqrt{s}$ center-of-mass energy. At $\sqrt{s}=7$ TeV and intermediate $p_\perp$, e.g. 2 GeV$/c$, gluons with $y\gtrsim3.5$ has $x>0.01$. They are called source gluons. Gluons with $y\lesssim1.0$ has $x<0.001$, which are descendants of source gluons and are called radiated gluons. In the central rapidity region, the small $x$ part of a colliding hadron is probed, while in the large rapidity region the large $x$ part is probed. A proton is seen as three valence quarks at large $x$ and dense gluons at small $x$. Therefore the physics in different rapidity regions is essentially different. Correlations of gluons of different rapidity regions can reflect correlations of different generations~\cite{Zhao-1,ZhangHY,Zhang-1,Zhao-2}.  

The experimental observable of ridge is per-trigger yield~\cite{CMS-2011-pp-7,ATLAS-2015-pp-13,CMS-2016-pp-13,CMS-2017-pp-13,CMS-2013-pPb-5.02,ALICE-2013-pPb-5.02,CMS-2013-pPb-PbPb,CMS-2011-pp-PbPb,CMS-2011-PbPb-2.76,CMS-2012-PbPb-2.76}, i.e. the number of particle pairs with pseudo-rapidity interval $\Delta\eta$ and azimuth interval $\Delta\phi$ divided by the number of trigger particles. In order to eliminate the influence of unrelated pairs, mixed events are usually constructed in experiments. Dividing the yield in real events by the yield in mixed events gives the final results reported in experiments. 

In the CGC calculation, we propose that the number of uncorrelated pairs can be exactly represented by the integral of the product of two real single-particle distributions within the acceptance, instead of the approximate normalization factor, which is given by simply assuming a boost invariant rapidity distribution~\cite{Dusling-2013-3}. As we know, the boost invariance of the rapidity distribution holds only at the central rapidity region~\cite{CGC-review}, but is violated beyond that region. The violation of boost invariance should influence correlations accordingly. Indeed, after this correction, the ridge correlations at long-range rapidity is well shown. The correlation rebound in p-p collisions at 7 TeV is reproduced and described by the CGC. 

It is also found that the correlation rebound is most obvious around the sum of the saturation momentum of the projectile and target. The rebound happens at even larger rapidity region for higher colliding energies. These features of the correlation rebound are closely related to the mechanism of CGC.

This paper is organized as follows. In section II we formulate an exact calculation of the normalization factor of ridge correlations based on real rapidity distributions of CGC. Results of corrected correlations are presented. In section III the dependence of the correlation rebound on $p_{\perp}$ and $\sqrt{s}$ is systematically studied and the origin is discussed. Section IV is a summary.  

\section{Long-range ridge correlations from CGC}

The per-trigger yield is defined as
\begin{equation}
\frac{1}{N_{\rm Trig}} \frac{\mathrm{d}^2 N^{\rm pair}}{\mathrm{d}\Delta y\mathrm{d}\Delta\phi}.
\end{equation}
It counts the number of particle pairs with rapidity separation $\Delta y$ and azimuthal angle separation $\Delta\phi$, divided by the number of trigger particles. 

In experiments, the number of uncorrelated pairs is estimated by the sample of mixed events. Particles of a mixed event are drawn randomly from different original events. For a large enough number of original events, in a single mixed event, the probability of having two particles from the same original event is close to zero.  The particles in a mixed event are almost independent~\cite{Gazd}.  

The per-trigger yield obtained from the original events and the mixed events are denoted as
\begin{eqnarray}
S(\Delta y,\Delta\phi)=\frac{1}{N_{\rm Trig}} \frac{\mathrm{d}^2 N^{\rm same}}{\mathrm{d}\Delta y\mathrm{d}\Delta\phi},\\
B(\Delta y,\Delta\phi)=\frac{1}{N_{\rm Trig}} \frac{\mathrm{d}^2 N^{\rm mixed}}{\mathrm{d}\Delta y\mathrm{d}\Delta\phi},
\end{eqnarray}  
respectively. The functions $S(\Delta y,\Delta\phi)$ and $B(\Delta y,\Delta\phi)$ are called the signal and background distributions, respectively. 
The final yield is normalized as~\cite{CMS-2011-pp-7,ATLAS-2015-pp-13,CMS-2016-pp-13,CMS-2017-pp-13,CMS-2013-pPb-5.02,ALICE-2013-pPb-5.02,CMS-2013-pPb-PbPb,CMS-2011-pp-PbPb,CMS-2011-PbPb-2.76,CMS-2012-PbPb-2.76}
\begin{equation}
Y(\Delta y,\Delta\phi)=B(0,0)\frac{S(\Delta y,\Delta\phi)}{B(\Delta y,\Delta\phi)}.
\end{equation}

Detector effects, such as tracking inefficiency, largely cancel in the same-event to mixed-event ratio. The factor $B(0,0)$ is the value of $B(\Delta y,\Delta\phi)$ at $\Delta y=0$ and $\Delta\phi=0$, representing the mixed-event associated yield for both particles of the pair going in the same direction. In this case, the two particles have the maximum pair acceptance and the normalization facor $B(\Delta y,\Delta\phi)/B(0,0)$ equals one.

The signal distribution of the per-trigger yield in CGC is expressed~~\cite{Dusling-2013-3,Dusling-2013-1,Dusling-2013-2} as
\begin{eqnarray}
S(\Delta y,\Delta\phi)=\frac{1}{N_{\rm Trig}} \frac{\mathrm{d}^2 N^{\rm assoc}}{\mathrm{d}\Delta y\mathrm{d}\Delta\phi},
\end{eqnarray}
where
\begin{widetext}
\begin{eqnarray}
\frac{\mathrm{d}^{2}N^{\rm{assoc}}}{\mathrm{d}\Delta y \mathrm{d}\Delta\phi} &=&\int_{y^{\rm min}}^{y^{\rm max}} \mathrm{d}y_{\rm p}\int_{y^{\rm min}}^{y^{\rm max}} \mathrm{d}y_{\rm q} \delta(y_{\rm q}-y_{\rm p}-\Delta y)\int_0^{2\pi} \mathrm{d}\phi_{\rm p} \int_0^{2\pi} \mathrm{d}\phi_{\rm q}  \delta(\phi_{\rm q}-\phi_{\rm p}-\Delta \phi) \nonumber\\ &&\times\int^{p^{\rm{max}}_\bot}_{ p^{\rm{min}}_\bot}\frac{\mathrm{d}p^{2}_\bot}{2}\int^{q^{\rm{max}}_\bot}_{q^{\rm{min}}_\bot}\frac{\mathrm{d}q^{2}_\bot}{2}\frac{\mathrm{d}N^{\rm{corr}}_{\mathrm{2}}}{{\mathrm{d}^{2}\boldsymbol{p}_\bot \mathrm{d}y_{\rm p}}{\mathrm{d}^{2}\boldsymbol{q}_\bot \mathrm{d}y_{\rm q}}}.
\end{eqnarray}\end{widetext}
The labels ``p" and ``q" denote the two particles in the pair, conventionally referred to as ``trigger" and ``associated" particles, respectively. The $\delta$ function is used to restrict the phase space interval to a given $\Delta y$ and $\Delta\phi$. The integrand function $\frac{\mathrm{d}N^{\rm{corr}}_{\mathrm{2}}}{{\mathrm{d}^{2}\boldsymbol{p}_\bot \mathrm{d}y_{\rm p}}{\mathrm{d}^{2}\boldsymbol{q}_\bot \mathrm{d}y_{\rm q}}}$ is equal to the two-particle production minus the product of two single-particle productions, i.e.
\begin{widetext}
\begin{equation}
\frac{\mathrm{d}N^{\rm{corr}}_{\mathrm{2}}}{{\mathrm{d}^{2}\boldsymbol{p}_\bot \mathrm{d}y_{\rm p}}{\mathrm{d}^{2}\boldsymbol{q}_\bot \mathrm{d}y_{\rm q}}}=\frac{\mathrm{d}N_{\mathrm{2}}}{{\mathrm{d}^{2}\boldsymbol{p}_\bot \mathrm{d}y_{\rm p}}{\mathrm{d}^{2}\boldsymbol{q}_\bot \mathrm{d}y_{\rm q}}}-\frac{\mathrm{d}N_1}{\mathrm{d}^{2}\boldsymbol{p}_\bot \mathrm{d}y_{\rm p}}\frac{\mathrm{d}N_1}{\mathrm{d}^{2}\boldsymbol{q}_\bot \mathrm{d}y_{\rm q}}.
\end{equation} 
\end{widetext}

The background in Eq. (4) represents yield of uncorrelated pairs. The counterpart in theoretical calculations should be integrals of the product of two single-particle productions, i.e.  
\begin{eqnarray}
B(\Delta y,\Delta\phi)=\frac{1}{N_{\rm Trig}} \frac{\mathrm{d}^2 N^{\rm uncorr}}{\mathrm{d}\Delta y\mathrm{d}\Delta\phi},
\end{eqnarray}
with
\begin{widetext}
\begin{eqnarray}
\frac{\mathrm{d}^{2}N^{\rm{uncorr}}}{\mathrm{d}\Delta y \mathrm{d}\Delta\phi} &=&\int_{y^{\rm min}}^{y^{\rm max}} \mathrm{d}y_{\rm p}\int_{y^{\rm min}}^{y^{\rm max}} \mathrm{d}y_{\rm q}\delta(y_{\rm q}-y_{\rm p}-\Delta y)\int_0^{2\pi} \mathrm{d}\phi_{\rm p} \int_0^{2\pi} \mathrm{d}\phi_{\rm q}  \delta(\phi_{\rm q}-\phi_{\rm p}-\Delta \phi) \nonumber\\ &&\times\int^{p^{\rm{max}}_\bot}_{p^{\rm{min}}_\bot}\frac{\mathrm{d}p^{2}_\bot}{2}\int^{q^{\rm{max}}_\bot}_{q^{\rm{min}}_\bot}\frac{\mathrm{d}q^{2}_\bot}{2}\frac{\mathrm{d}N_1}{\mathrm{d}^{2}\boldsymbol{p}_\bot \mathrm{d}y_{\rm p}}\frac{\mathrm{d}N_1}{\mathrm{d}^{2}\boldsymbol{q}_\bot \mathrm{d}y_{\rm q}}.
\end{eqnarray}\end{widetext}

The integration in Eq. (10) depends on the shape of the single-particle distribution and the acceptance. The CMS, ALICE and ATLAS experiments at the LHC have a full azimuthal coverage but a limited rapidity acceptance. When the single-particle azimuthal distribution is uniform and the integral range is 0 to $2\pi$, the background distribution does not depend on $\Delta\phi$. 

On contrary, the background distribution depends on $\Delta y$ due to the limited rapidity acceptance. When the rapidity distribution is boost invariant, the normalization factor is
\begin{eqnarray}
\frac{B(\Delta y,\Delta\phi)}{B(0,0)}=1-\frac{|\Delta y|}{y^{\rm max}-y^{\rm min}},
\end{eqnarray}
i.e. Eq. (A.4) in ref.~\cite{Dusling-2013-3}.


As we know, boost invariance of glasma fields only holds approximately within small $x$ region in the CGC framework. It is interesting to see how an exact calculation of the normalization factor affects the correlations. In the following, the background based on boost invariance, i.e. Eq. (11), is denoted by $B_{\rm 1}$.  The background based on real single-particle distributions  within the CGC is denoted by $B_{\rm 2}$. The signal distribution $S$ normalized by $B_1$, $B_2$ results in $Y_{\rm 1}$ and $Y_{\rm 2}$, respectively.

The quantum evolution with rapidity is described by the rcBK equation~\cite{rcBK-1,rcBK-2,rcBK-3}. By solving the rcBK equation at a given initial condition, the unintegrated gluon distribution can be obtained and the two- and single-gluon productions are available. To avoid repetition, the formulae of the double-gluon production and sing-gluon production presented in refs.~\cite{Dusling-2013-3,Dusling-2013-1,Dusling-2013-2,Zhao-1,ZhangHY,Zhang-1} are not shown here.  Completing the integrals in Eqs.~(7) and (10) with the transverse momentum range $1\leq p_{\perp}(q_{\perp})\leq 3~$GeV$/c$ and the rapidity range $-0.9\leq y_{\rm p}(y_{\rm q})\leq 0.9$ and $-2.4\leq y_{\rm p}(y_{\rm q})\leq 2.4$, $Y_{\rm 1}$ and $Y_{\rm 2}$ are obtained and shown in Fig.~1. The purpose of using two different rapidity windows are to distinguish the contributions of different $x$ components. $Y_{\rm w}$ in the figure is short for rapidity window.

\begin{figure*}[ht]
\begin{center}
\includegraphics[scale=0.34]{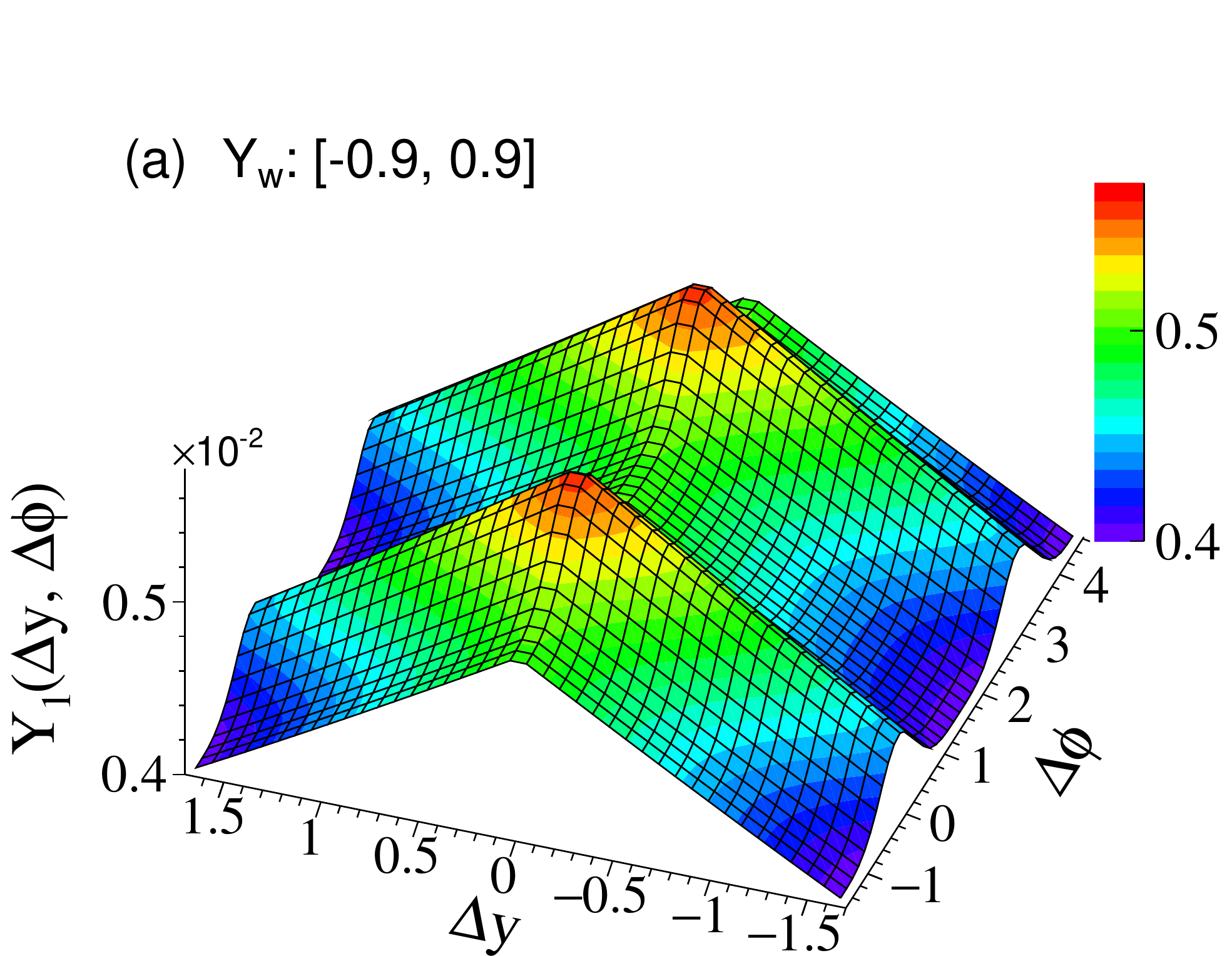}
\includegraphics[scale=0.34]{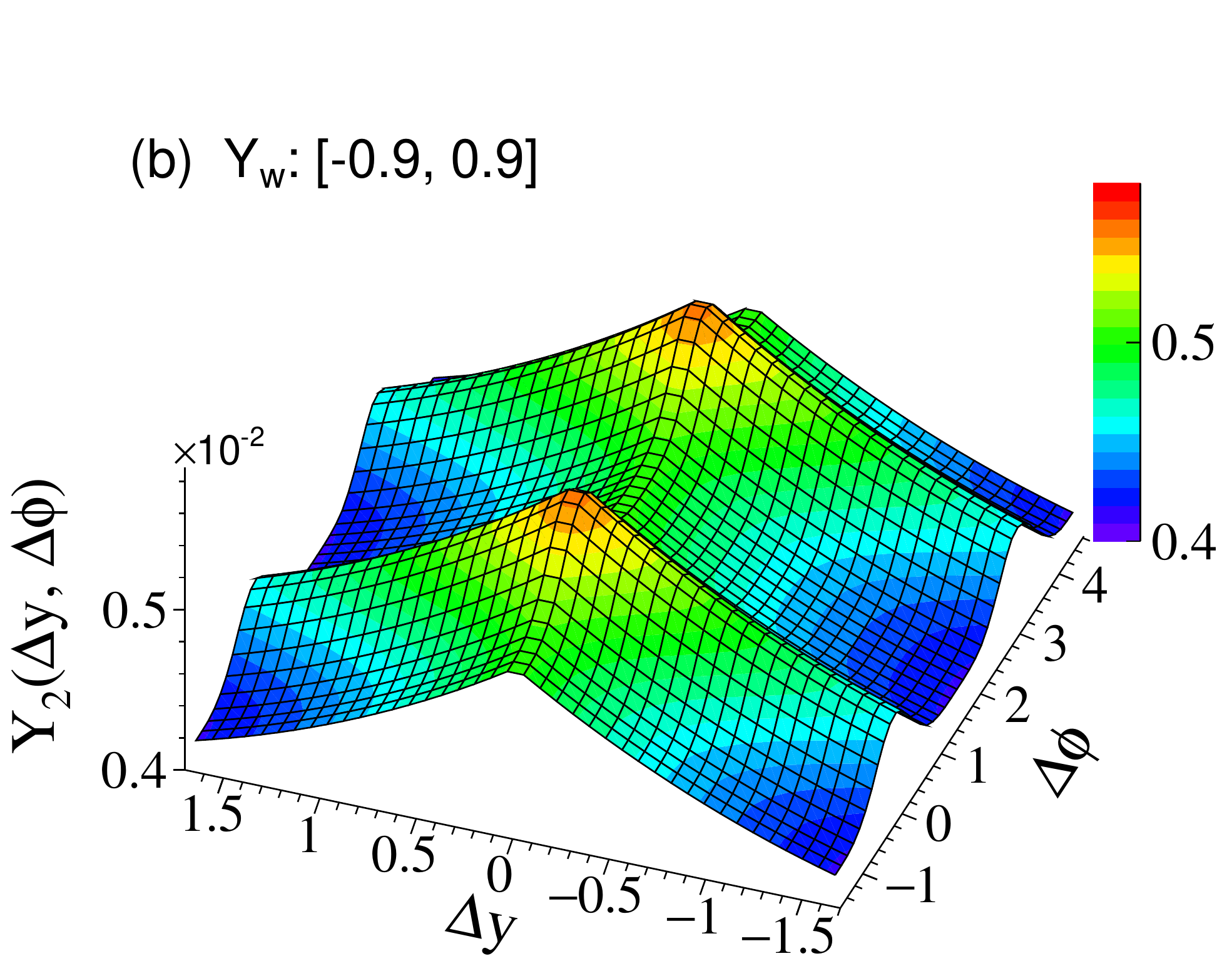}
\includegraphics[scale=0.34]{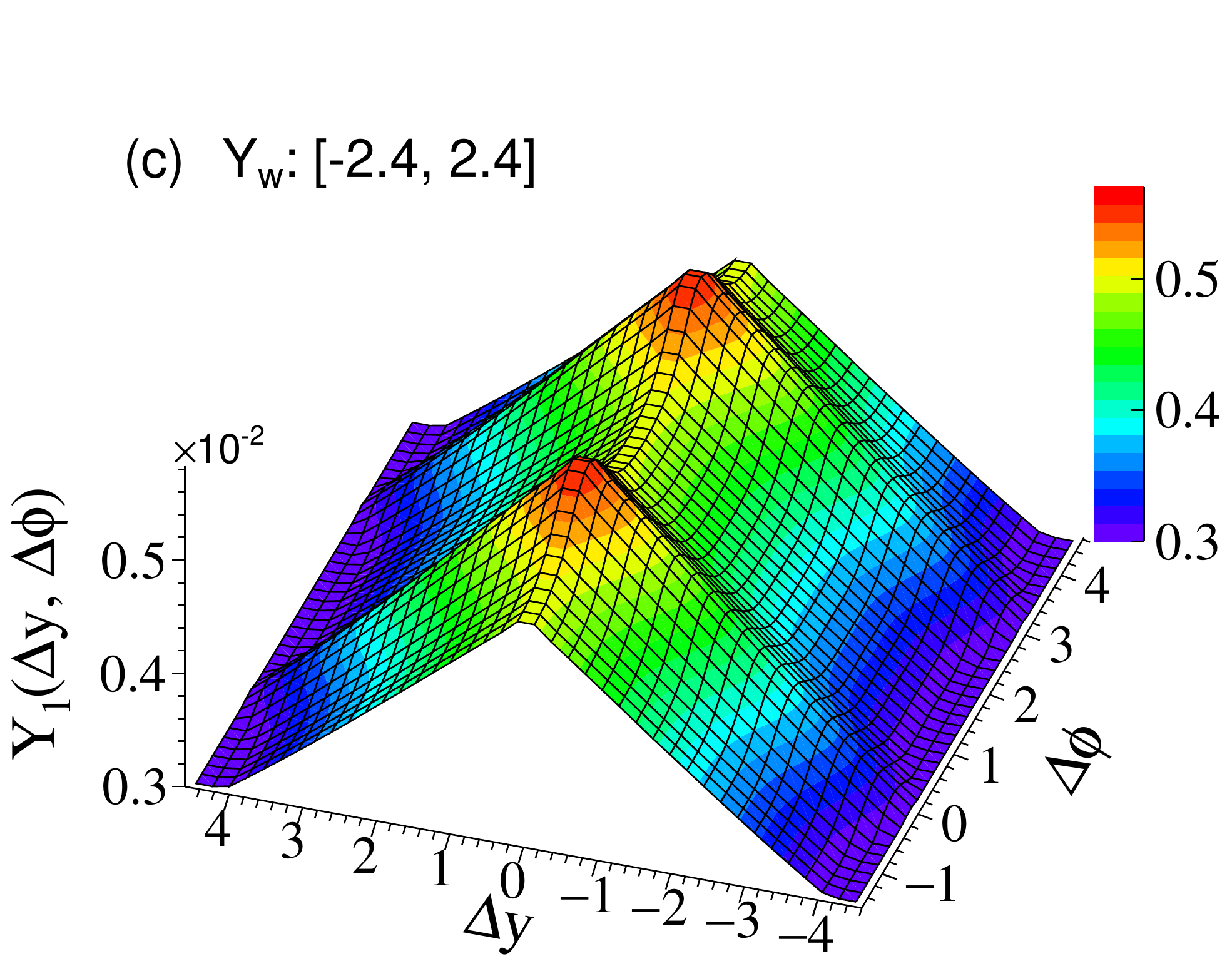}
\includegraphics[scale=0.34]{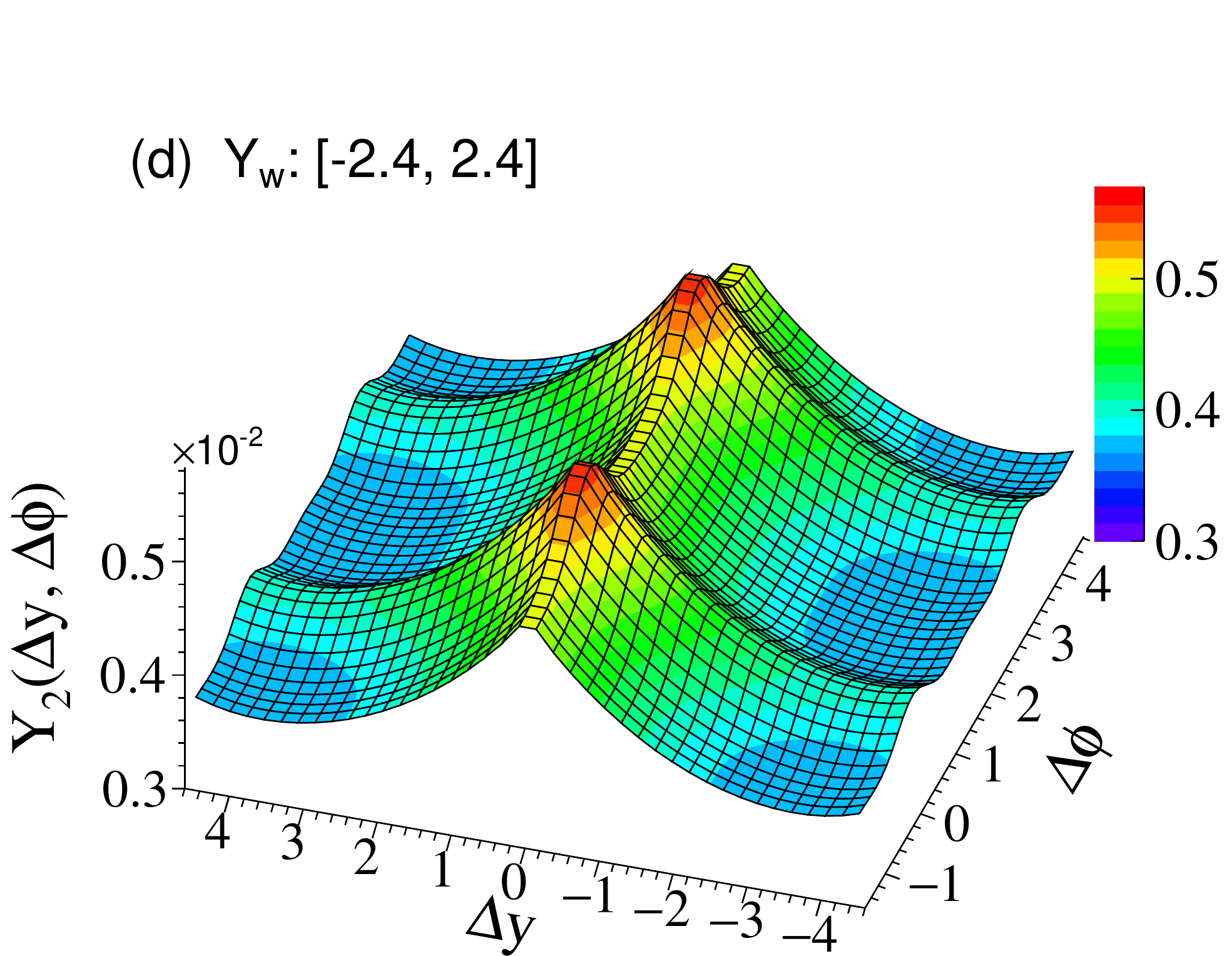}
\end{center}
\caption{The per-trigger yield in the $\Delta y$-$\Delta\phi$ plane  for 7 TeV pp collisions with transverse momentum integrated within $1\leq p_{\perp}(q_{\perp})\leq 3$~GeV$/c$ and with rapidity integrated in $-0.9\leq y_{\rm p}(y_{\rm q})\leq 0.9$ (the upper panels) and $-2.4\leq y_{\rm p}(y_{\rm q})\leq 2.4$ (the lower panels). The two columns present quantities $Y_{1}$ and $Y_{2}$, respectively. }
\end{figure*}

As Figs.~1(a) and 1(b) show, at the rapidity window of $[-0.9, 0.9]$ (the ALICE acceptance), $Y_{\rm 1}$ and $Y_{\rm 2}$ have similar structures. In the $\Delta\phi$ direction, they both have two peaks of equal height at $\Delta\phi=0$ and $\pi$. The two peaks are called azimuthal collimation which is intrinsic to glasma dynamics~\cite{Dusling-2013-1,Dusling-2013-2,Dusling-2013-3}. It contributes to the well-known collectivity in small systems. In the $\Delta y$ direction, $Y_{\rm 1}$ and $Y_{\rm 2}$ both show a downward trend as $|\Delta y|$ increases. It was stated that glasma graphs have significant short range rapidity correlations~\cite{Zhao-2}. Due to the short range rapidity correlations, the longitudinal structure of the two dimensional distributions is not as flat as the ALICE data~\cite{ALICE-2013-pPb-5.02}. So at the rapidity window of $[-0.9, 0.9]$ the results of two dimensional distributions from CGC are not directly comparable with data.

Since boost invariance holds approximately within small $x$ region in the CGC framework, it is undersdandable that two normalization schemes have few differences within the central rapidity region. 

At the rapidity window of $[-2.4, 2.4]$ (the CMS acceptance), $Y_{\rm 1}$ and $Y_{\rm 2}$ are shown in Figs.~1(c) and 1(d), respectively. In Fig.~1(c) correlations at long-range rapidity ($|\Delta y|\gtrsim2$) show bumps at $\Delta\phi=0$ and $\pi$ which contribute to the ridge yield. Integrating the two dimensional distribution within $2<|\Delta y|<4$ produces the ridge yield as a function of $\Delta\phi$, i.e. $\frac{\mathrm{d}N}{\mathrm{d}\Delta\phi}$, which has been shown to be consistent with the CMS data~\cite{Dusling-2013-1,Dusling-2013-2,Dusling-2013-3}. 

Of particular interest in this study is the ridge yield as a function of $\Delta y$, which has not been demonstrated within the CGC framework~\cite{Dusling-2013-3,Dusling-2013-1,Dusling-2013-2}.  The near-side yields as a function of $\Delta\eta$ from the CMS data presents a rebound at $|\Delta\eta|\approx4$ after a plateau within $2<|\Delta\eta|<3.6$ (see Fig. 2 in Ref.~\cite{CMS-2011-pp-7}). By using the background $B_1$, the ridge yield as a function of $\Delta y$, as Fig.~1(c) shows, does not agree with the data, i.e. the plateau and the rebound in the rapidity direction are not reproduced. 

However, the trend of $Y_{\rm 2}$ is qualitatively different from $Y_{\rm 1}$, as Fig.~1(d) shows. The difference is mainly in the rapidity direction. Correlations in Fig.~1(c) decrease with $|\Delta y|$ monotonously. Correlations in Fig.~1(d) first fall and then rise with $|\Delta y|$, indicating a clearer signal of long-range rapidity correlations. The structure within $2<|\Delta y|<4.4$ is qualitatively consistent with the CMS data~\cite{CMS-2011-pp-7}.

In order to compare $Y_{1}$ and $Y_{2}$ more clearly, a projection to $\Delta y$ axis is made and shown in Figs.~2(a) and 2(b). In Fig.~2(a), at the rapidity window of $[-0.9, 0.9]$, the red curve (representing $Y_2$) is almost coincident with the black curve (representing $Y_1$), only having visible differences at $|\Delta y|>1$. This means that the two backgrounds are approximately equal and the $Y$ quantity is also approximately equal when the rapidity window is within the central rapidity region. This is because the boost invariance holds approximately in the central rapidity region, and $Y_{2}$ almost reduces to $Y_{1}$.

As the rapidity window increases to $[-2.4, 2.4]$, the red curve in Fig.~2(b) deviates the black curve significantly at long-range rapidity of $|\Delta y|>2$. The plateau and the rebound in the rapidity direction in CMS data~\cite{CMS-2011-pp-7} are reproduced by the red curve in Fig.~2(b). 
\begin{figure*}[htbp]
\begin{centering}
\begin{tabular}{cc}
$\vcenter{\hbox{\includegraphics[scale=0.31]{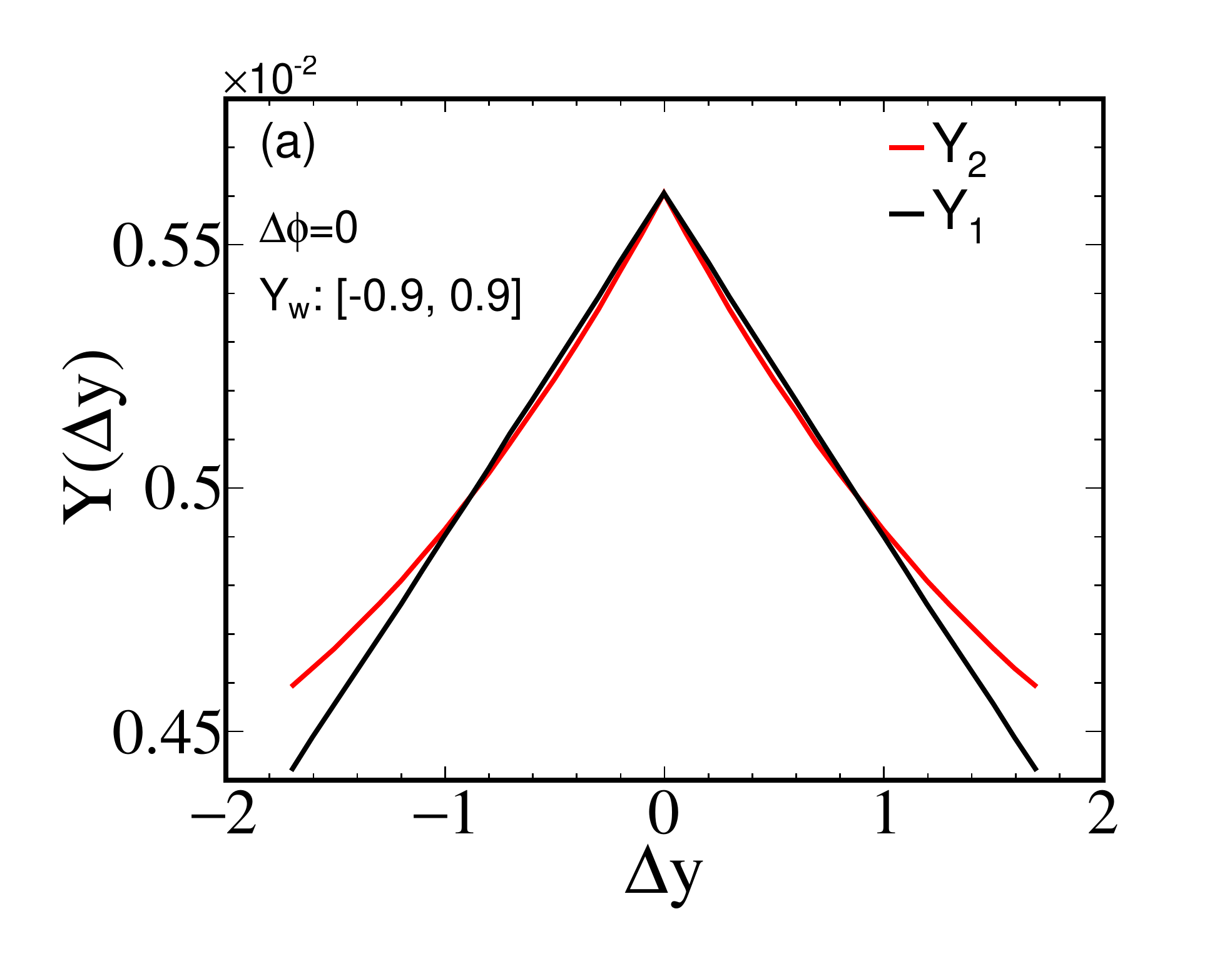}}}$
$\vcenter{\hbox{\includegraphics[scale=0.31]{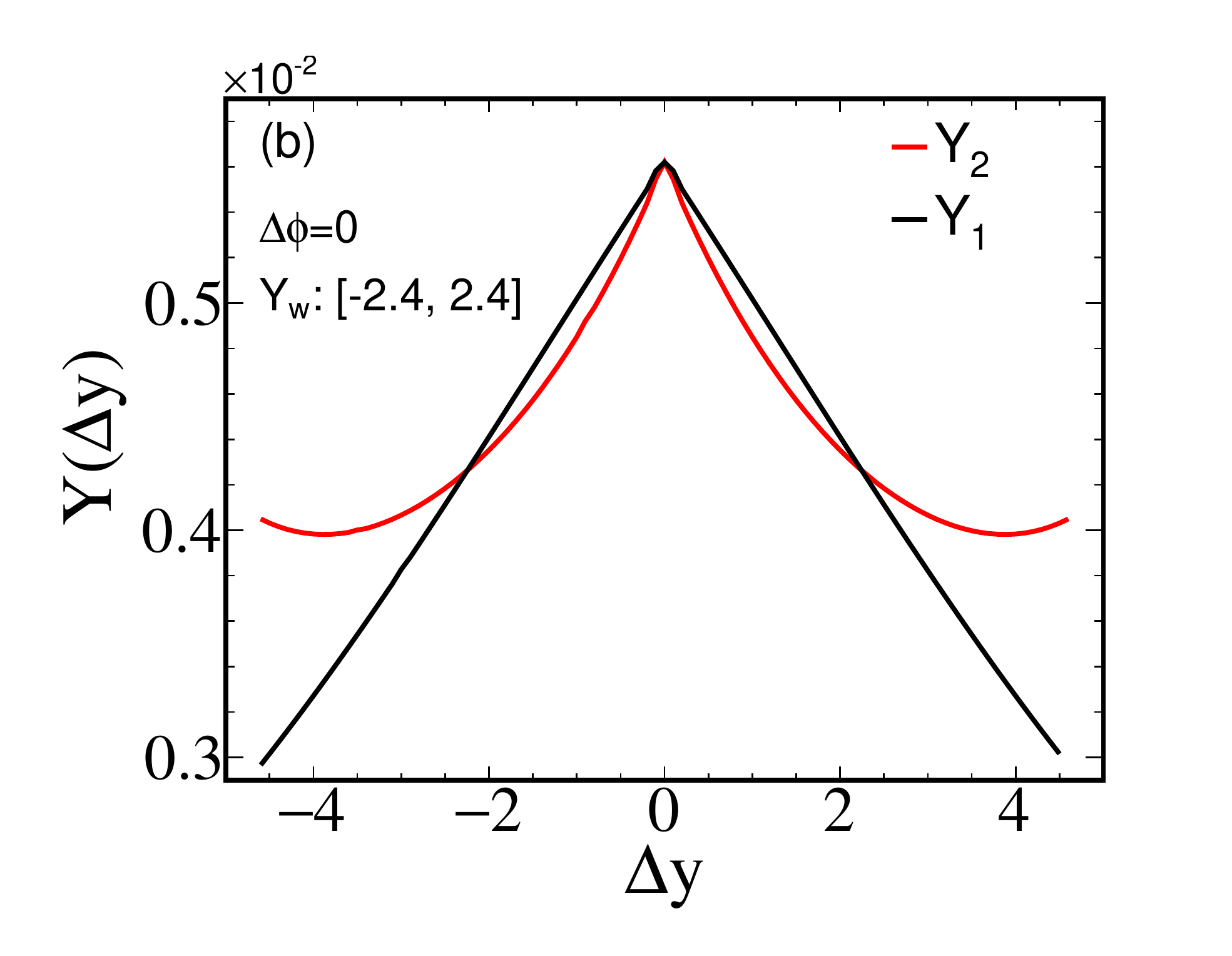}}}$ &
$\vcenter{\hbox{\includegraphics[scale=0.31]{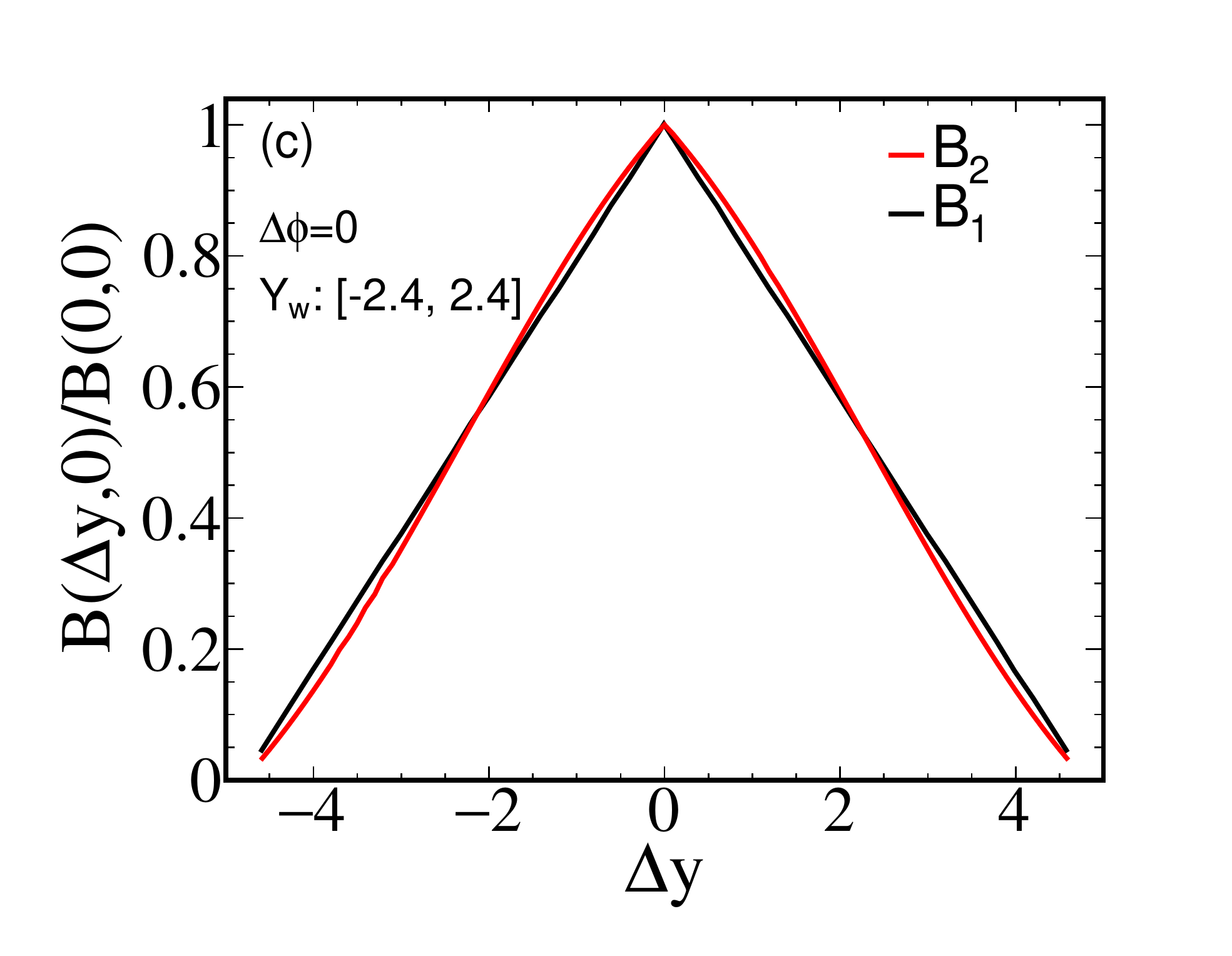}}}$
\end{tabular}
\par\end{centering}
\caption{The per-trigger yield as a function of $\Delta y$ at fixed $\Delta\phi=0$ for the rapidity windows of $[-0.9, 0.9]$ (a) and $[-2.4, 2.4]$ (b) . The black and the red curves are $Y_{1}$ and $Y_{2}$, respectively.  (c) The difference of $B_1$ and $B_2$ at the rapidity window of $[-2.4, 2.4]$. }
\end{figure*}

The trend of $Y_1$ as a function of $\Delta y$ does not show any rebound at all and thus does not match well with data. $Y_2$ (the red curve) demonstrates a correlation rebound at long-range rapidity of $|\Delta y|\approx4$. This may be the first time that the rapidity correlation calculated within the CGC framework agrees with the experimental data in trend. The better agreement is due to the correction of the normalization scheme of the ridge yield.

The qualitative differences between $Y_{1}$ and $Y_{2}$ originate from the differences in normalization factor $B_{1}$ and $B_{2}$, which are shown in Fig.~2(c). The black curve (representing $B_{1}$) and the red curve (representing $B_{2}$) nearly touch each other except minor differences. $B_2$ is a little larger than $B_1$ at $|\Delta y|<2$ and slightly smaller at $|\Delta y|>2$. The reason for that is the violation of the boost invariance of the rapidity distribution at the large-rapidity region.

As mentioned before, the normalization factor appears in the denominator of Eq. (5). Its tiny differences cause a qualitative change in $Y$ quantity. That is why an accurate calculation of the  normalization factor is highly important.

\section{The $p_\perp$ and $\sqrt{s}$ dependence of large-rapidity ridge like correlations}

In the following, it is interesting to see if and how the rebound of rapidity correlations changes with transverse momentum and colliding energy. 

The ridge yield $Y_2$ at $\sqrt{s}=7$ TeV and 13 TeV are shown in Fig. 3. The color codes of the surface plots are set to be the same for the sake of comparison. 

The upper panels of Fig. 3 are for rapidity window of $[-0.9, 0.9]$. The red area in Fig.~3(b) seems to be larger than Fig.~3(a), indicating  stronger correlations at $\Delta y=0$ when $\sqrt{s}=13$ TeV. Except this, the trend along the rapidity direction is nearly identical at $\sqrt{s}=7$ and 13 TeV. When the rapidity window is narrow, i.e. within small $x$ region, the ridge correlations have not much dependences on colliding energy. 

The lower panels of Fig. 3 are for rapidity window of $[-2.4, 2.4]$. As in the previous case, results of 7 TeV and 13 TeV does not show significant differences. However, compared with rapidity window of $[-0.9, 0.9]$, rebounds at large $|\Delta y|$ are significant in this case. In a narrow rapidity window, only correlations between small $x$ gluons contribute. In a wide rapidity window correlations between small $x$ gluons and large $x$ gluons contribute. Therefore, the rebound at large $|\Delta y|$ results from correlations of source gluons and radiated gluons.

A projection to $\Delta y$ axis is made and shown in Fig.~4(a). The solid lines are the projection of Figs.~3(c) and 3(d), whose transverse momentum interval is $[1, 3]$ GeV$/c$. The black solid curve (representing 13 TeV) is qualitatively different from the red solid curve (representing 7 TeV). The red solid curve first falls and then rises with $|\Delta y|$, showing a plateau and a rebound at long range rapidities. The black solid curve decreases with $|\Delta y|$ monotonically. Whether there is a rebound in the rapidity correlations is the main difference between the two energies.  

\begin{figure*}[ht]
\begin{center}
\includegraphics[scale=0.34]{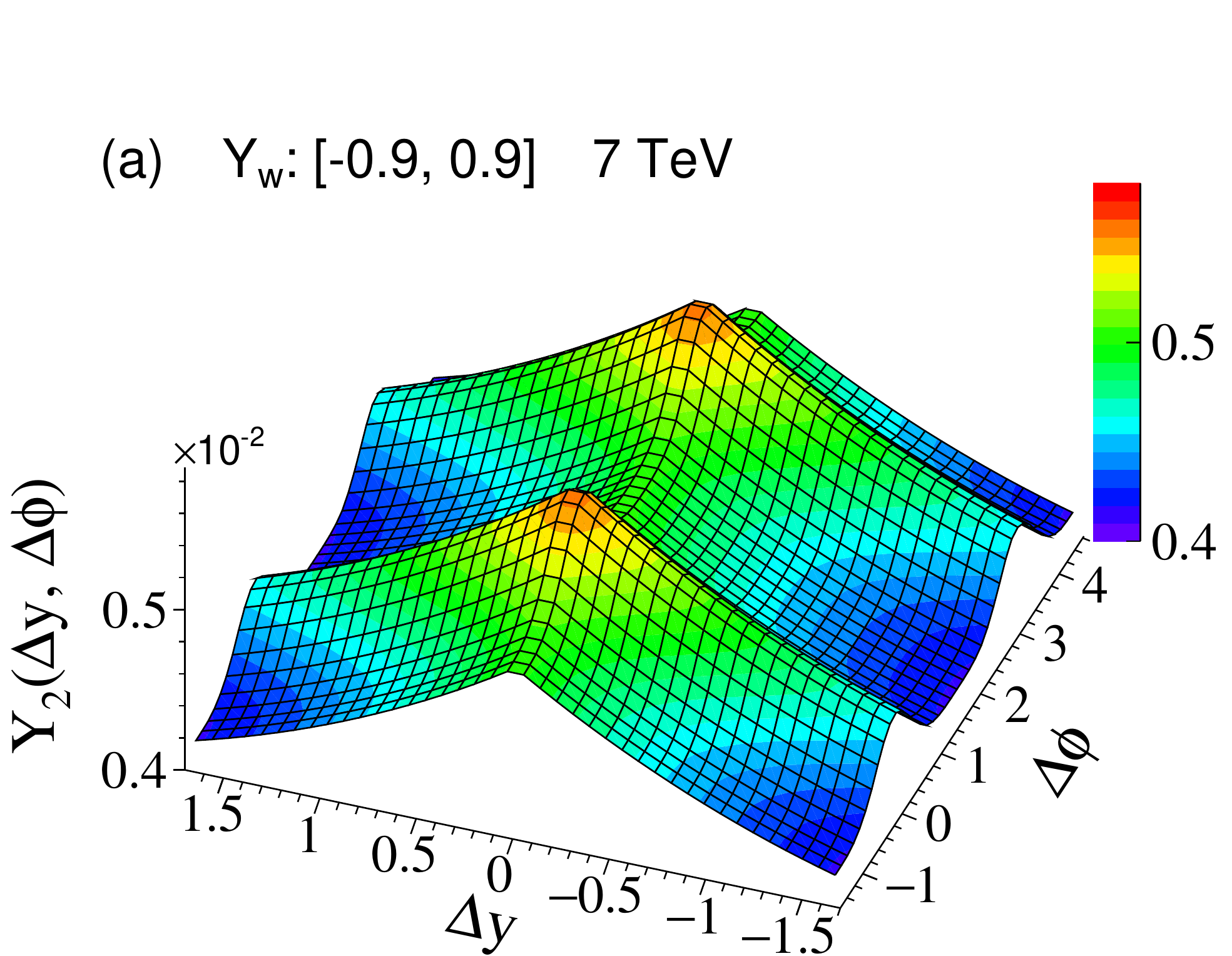}
\includegraphics[scale=0.34]{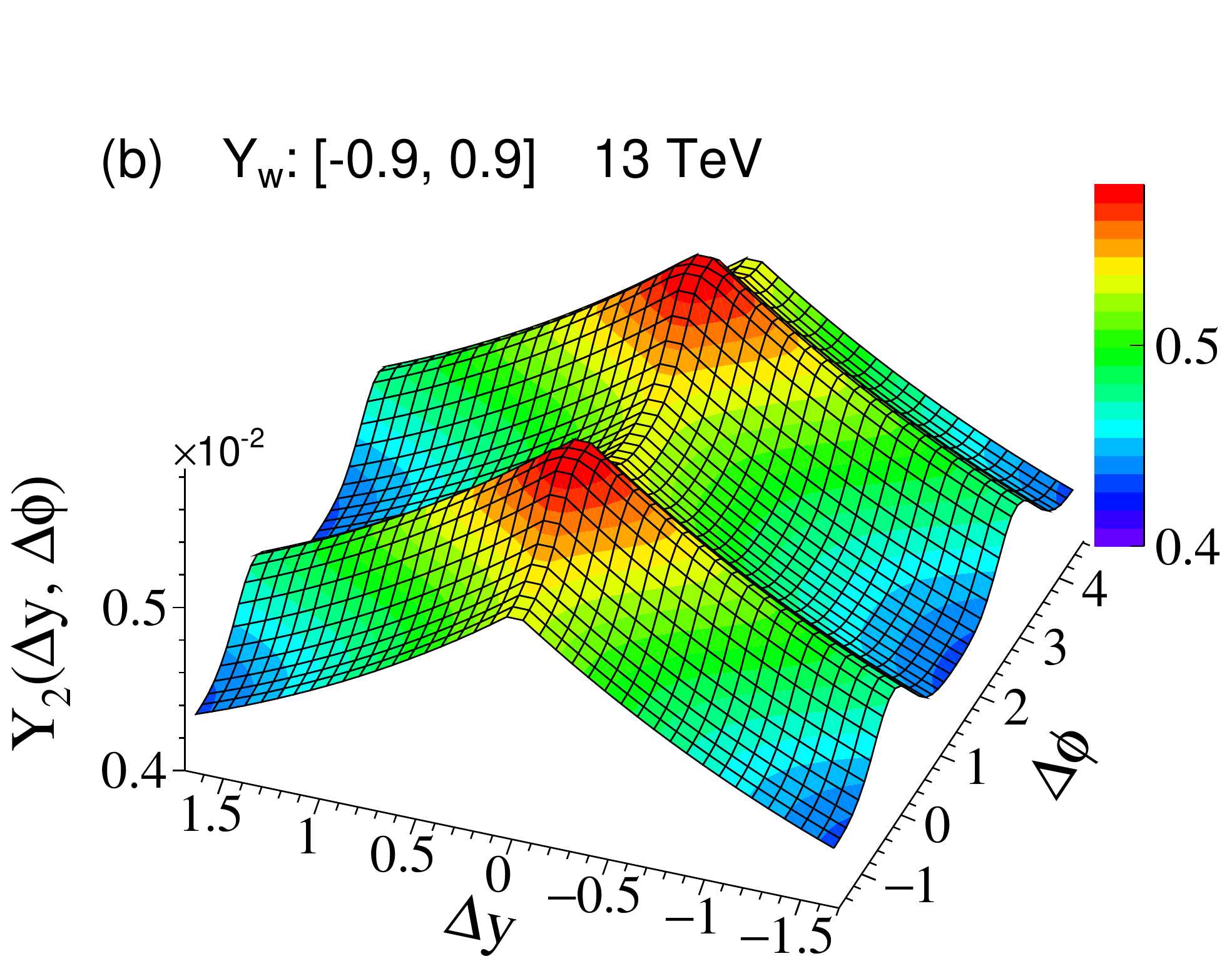}
\includegraphics[scale=0.34]{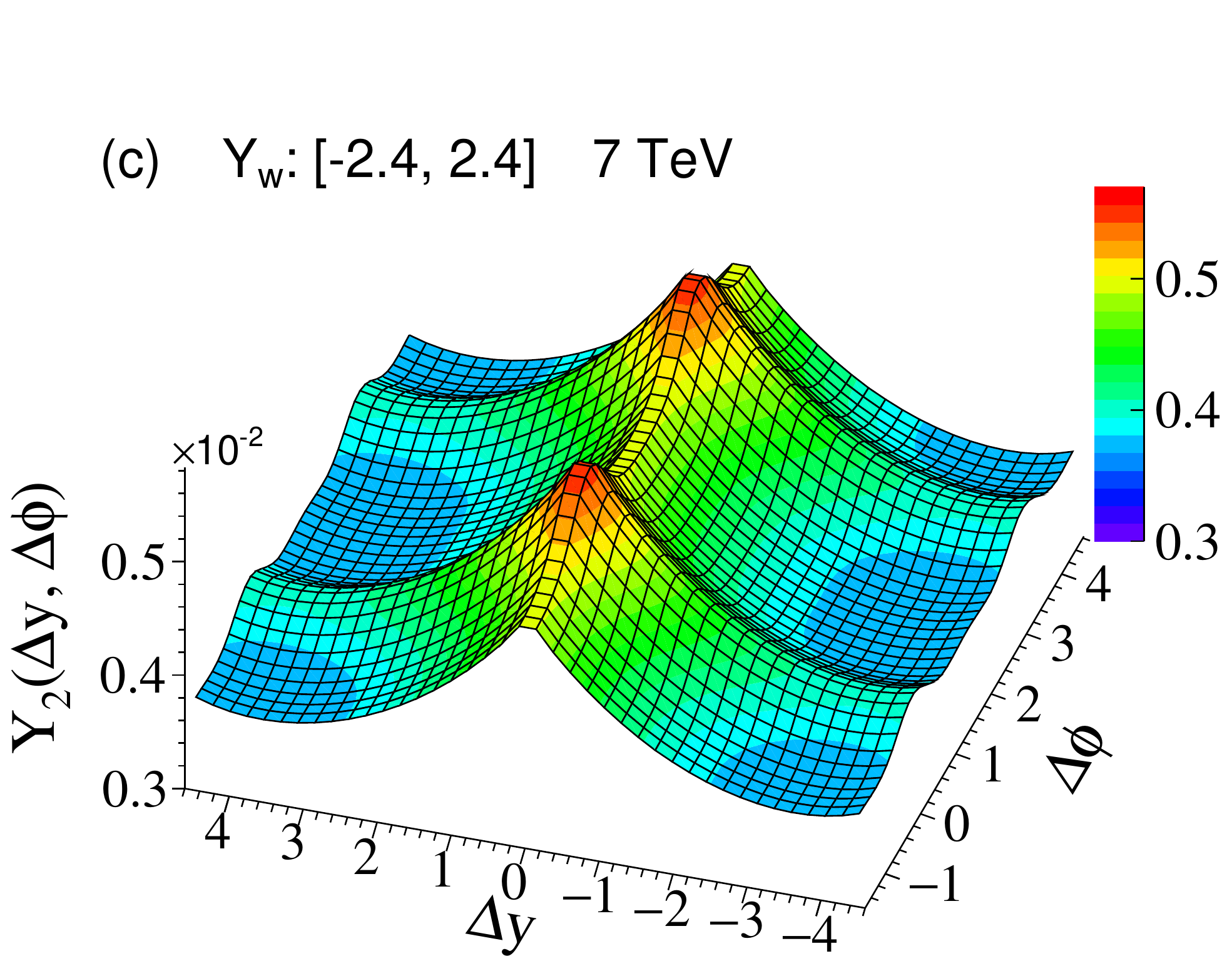}
\includegraphics[scale=0.34]{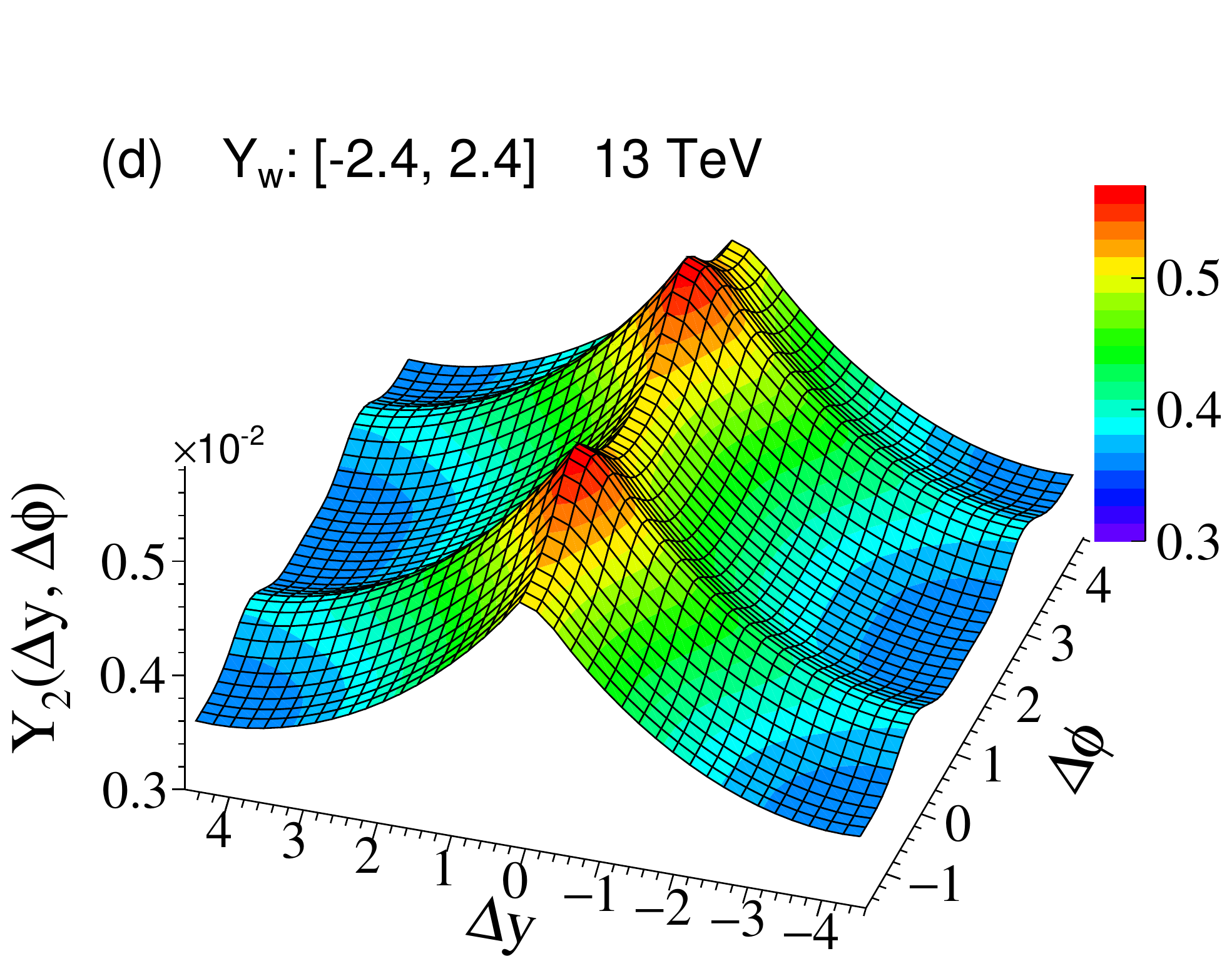}
\end{center}
\caption{The per-trigger yield $Y_{2}$ for pp collisions with transverse momentum integrated within $1\leq p_{\perp}(q_{\perp})\leq 3$~GeV$/c$ and with rapidity integrated in $-0.9\leq y_{\rm p}(y_{\rm q})\leq 0.9$ (the upper panels) and $-2.4\leq y_{\rm p}(y_{\rm q})\leq 2.4$ (the lower panels). The two columns present per-trigger yield of 7 TeV and 13 TeV, respectively. }
\end{figure*}

In order to study the transverse momentum dependence, the transverse momentum interval $[1, 3]$ GeV$/c$ is divided into two intervals, i.e. $[1, 2]$ GeV$/c$ and $[2, 3]$ GeV$/c$. The rapidity correlations within $[1, 2]$ GeV$/c$ (dashed lines) completely reproduce the trend of the solid lines. The rebound of the red solid curve at $|\Delta y|\approx 4.0$ is dominated by the transverse momentum interval $[1, 2]$ GeV$/c$. The correlations at  $p_\perp\in[2, 3]$ GeV$/c$ (the dot-dashed line) does not show any rebound trends. It indicates that the rebound of rapidity correlations at $|\Delta y|\approx 4.0$ is most obvious at $p_\perp\sim Q_{\rm sA}+Q_{\rm sB}=2Q_{\rm sp}\approx1.8$ GeV$/c$, where $Q_{\rm sA(B)}$ denotes the saturation momentum of the projectile or target, and $Q_{\rm sp}\approx 0.9$ GeV$/c$ at 7 TeV. This is consistent with the existing experimental result that ridge yield gets the maximum within $[1, 2]$ GeV$/c$ of particle transverse momentum~\cite{CMS-2010-pp}.

This $p_\perp$ dependences of the ridge correlations can be well explained under the CGC framework. The correlation function is proportional to the correlated two-gluon production, i.e. Eq.~(8), which can be expressed by convolutions of four uGDs~\cite{Dusling-2013-3,Dusling-2013-1,Dusling-2013-2,Zhao-1,ZhangHY,Zhang-1}, e.g.
\begin{equation}
\Phi^2_{\rm A}(y_{\rm p},\boldsymbol{k}_\bot)\Phi_{\rm B}(y_{\rm p},\boldsymbol{p}_\bot-\boldsymbol{k}_\bot)\Phi_{\rm B}(y_{\rm q},\boldsymbol{q}_\bot-\boldsymbol{k}_\bot).
\end{equation}
Since uGD ($\Phi$) peaks at $Q_{\rm s}$, transverse momentum far from $Q_{\rm s}$ contributes little to the correlation. To make a significant contribution to the correlation function, 
\begin{equation}
|\bo{k}_\perp|\sim Q_{\mathrm{s}},\quad |\bo{p}_\perp-\bo{k}_\perp|\sim Q_{\mathrm{s}}\quad \text{and}\quad |\bo{q}_\perp-\bo{k}_\perp|\sim Q_{\mathrm{s}}
\end{equation}
are required simultaneously. It approximately requires that $|\bo{p}_\perp|\sim|\bo{q}_\perp|\sim 2Q_{\rm s}$.

As we know, CGC has a consistent description for different colliding systems. The only parameter is the saturation momentum $Q_{\rm s}$. $Q_{\rm s}$ is $x$ dependent. As Eq. (1) demonstrates, when $\sqrt{s}$ increases, the rapidity $y$ must increase to get the same $x$. We expect the rebound of rapidity correlations should appear at larger rapidities for higher colliding energies. One of the variables of per-trigger yield is $\Delta y$, which reflects the rapidity gap rather than the rapidity location. Correlations with rapidity location as an independent variable is the differential correlation function, i.e.
\begin{widetext}
\begin{eqnarray}
C(\boldsymbol{p}_\bot, y_{\rm p}; \boldsymbol{q}_\bot, y_{\rm q}) &=& \frac{\frac{\mathrm{d}N_2}{{\mathrm{d}^{2}\boldsymbol{p}_\bot \mathrm{d}y_{\rm p}}{\mathrm{d}^{2}\boldsymbol{q}_\bot \mathrm{d}y_{\rm q}}}}{\frac{\mathrm{d}N_1}{\mathrm{d}^{2}\boldsymbol{p}_\bot \mathrm{d}y_{\rm p}}\frac{\mathrm{d}N_1}{\mathrm{d}^{2}\boldsymbol{q}_\bot \mathrm{d}y_{\rm q}}}-1= \frac{\frac{\mathrm{d}N_{\mathrm{2}}^{\rm corr}}{{\mathrm{d}^{2}\boldsymbol{p}_\bot \mathrm{d}y_{\rm p}}{\mathrm{d}^{2}\boldsymbol{q}_\bot \mathrm{d}y_{\rm q}}}}{\frac{\mathrm{d}N_1}{\mathrm{d}^{2}\boldsymbol{p}_\bot \mathrm{d}y_{\rm p}}\frac{\mathrm{d}N_1}{\mathrm{d}^{2}\boldsymbol{q}_\bot \mathrm{d}y_{\rm q}}}.
\end{eqnarray}
\end{widetext}
Here $\frac{\mathrm{d}N_2}{{\mathrm{d}^{2}\boldsymbol{p}_\bot \mathrm{d}y_{\rm p}}{\mathrm{d}^{2}\boldsymbol{q}_\bot \mathrm{d}y_{\rm q}}}$ and $\frac{\mathrm{d}N_1}{\mathrm{d}^{2}\boldsymbol{p}_\bot \mathrm{d}y_{\rm p}}$ are the two-gluon production and the single-gluon production, respectively, the same with that in Eq. (8). 

$C(\boldsymbol{p}_\bot, y_{\rm p}; \boldsymbol{q}_\bot, y_{\rm q})$  as a function of $y_{\rm q}$, i.e. $\Delta y$, for $y_{\rm p}=0$, $p_\bot=q_\bot=1.5$ GeV$/c$, $\phi_{\rm p}=\phi_{\rm q}=0$ is shown in Fig. 4(b). The trigger particle has $y_{\rm p}=0$, being at the central rapidity region. The associated particle has $y_{\rm q}$. The solid, dashed and dot-dashed lines denote small $x$, middle $x$ and large $x$ regions of the associated particle, respectively. The peak around $\Delta y=0$ reflects correlations between radiated gluons. The peak at $\Delta y\approx 4.0$ reflects correlations between radiated gluons ($y_{\rm p}=0$) and source gluons ($y_{\rm q}\approx 4.0$).  So the rebound of rapidity correlations at $|\Delta y|\approx 4.0$ is caused by the strong correlations between the radiated gluons and source gluons. 

\begin{figure*}[htbp]
\begin{centering}
\begin{tabular}{cc}
$\vcenter{\hbox{\includegraphics[scale=0.38]{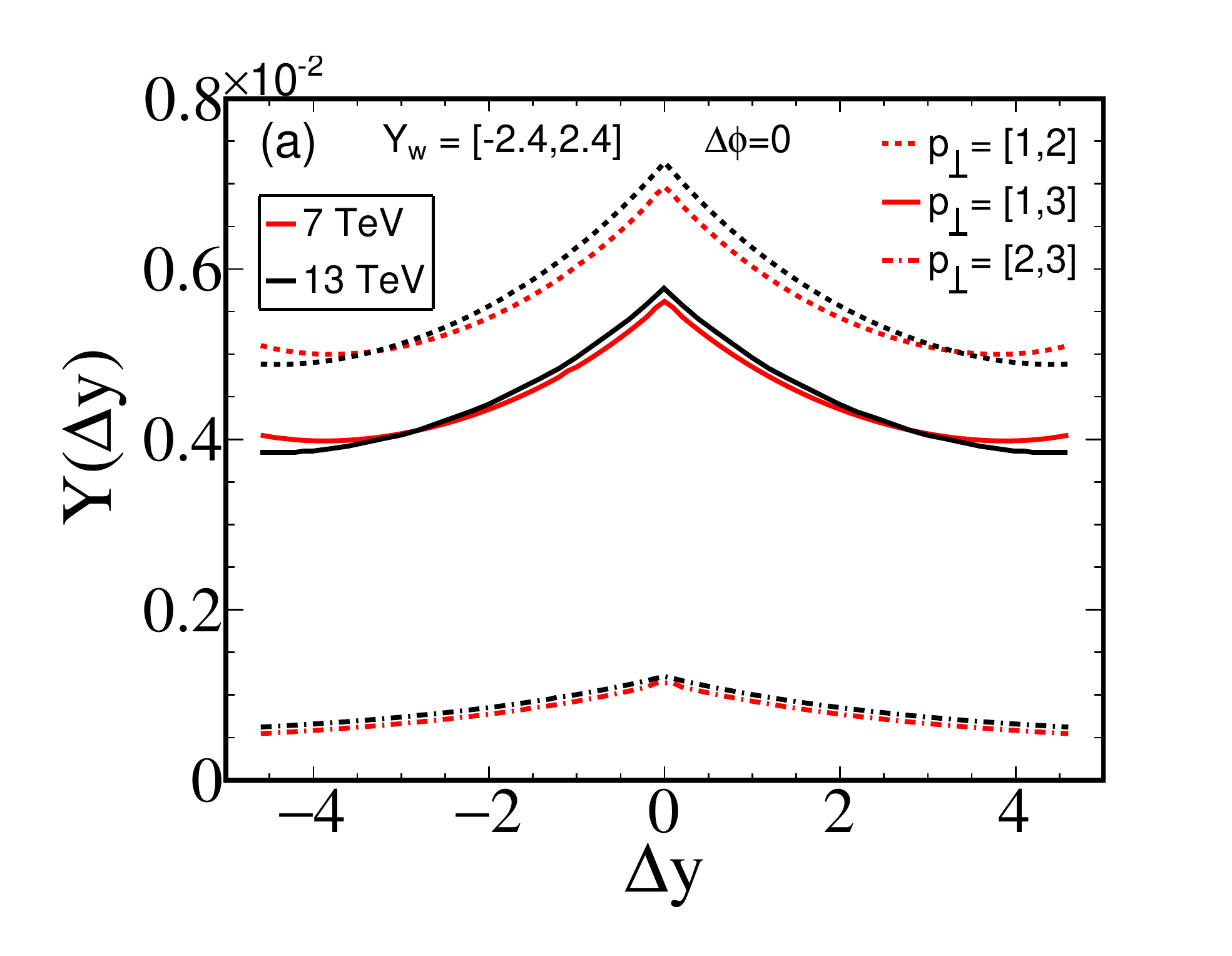}}}$ &
$\vcenter{\hbox{\includegraphics[scale=0.38]{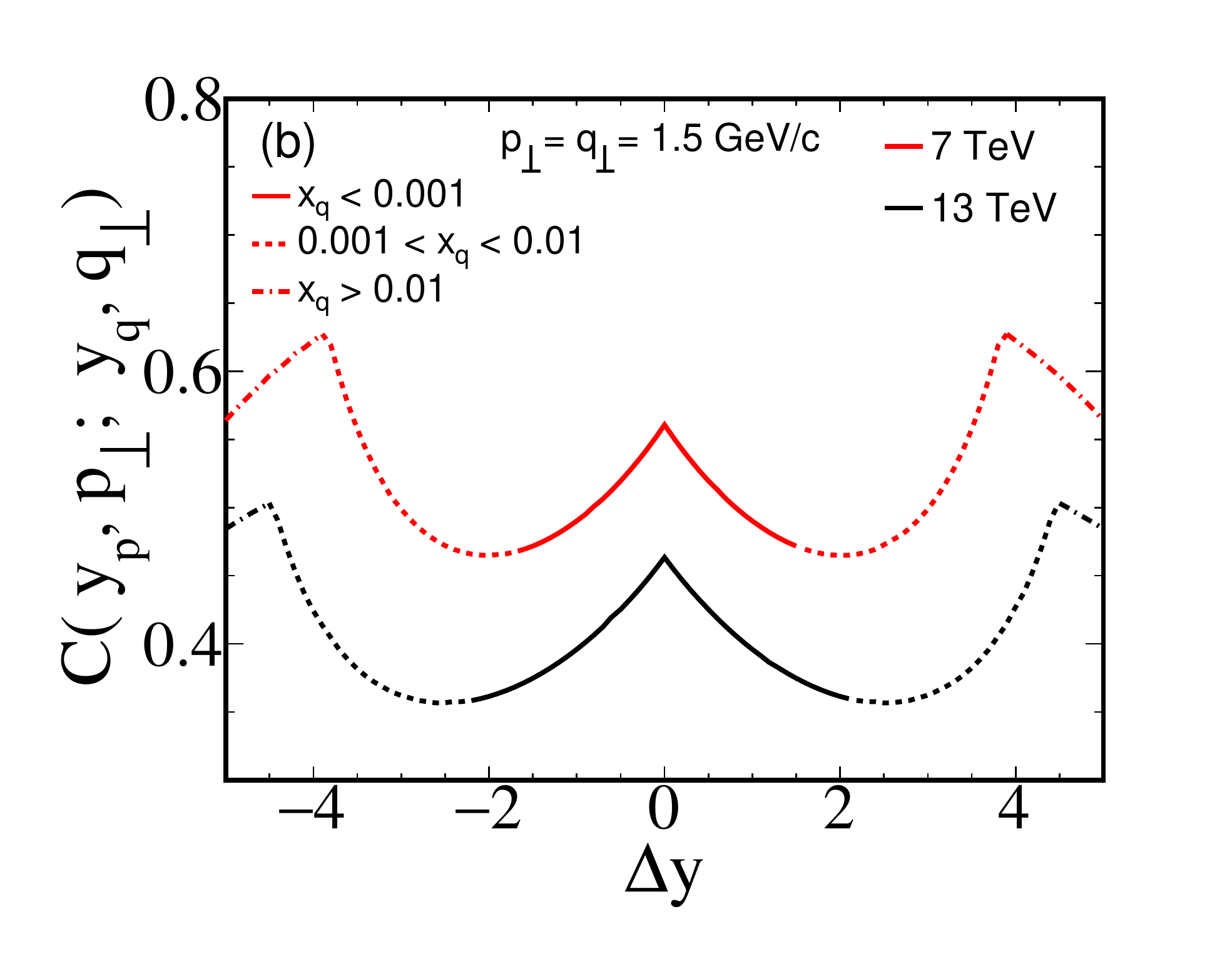}}}$
\end{tabular}
\par\end{centering}
\caption{(a) The per-trigger yield as a function of $\Delta y$ at $\Delta\phi=0$ for the rapidity window of $[-2.4, 2.4]$. The red and the black curves are for 7 TeV and 13 TeV, respectively. The solid line, dashed line and dot-dashed line represent ridge yield of transverse momentum interval $[1, 3]$, $[1, 2]$ and $[2, 3]$ GeV$/c$, respectively. (b) Differential correlation function as a function of $\Delta y$ at $p_{\perp}=q_{\perp}=1.5$ GeV$/c$ and $\Delta\phi=0$. The red and black curves are for 7 TeV and 13 TeV, respectively. }
\end{figure*}

The red (for 7 TeV) and black (for 13 TeV) curves have similar large-rapidity ridge correlations. The connection point of the dashed line and the dot-dashed line represents $x_{\rm q}=0.01$. Its position is $y_{\rm q}\approx3.84$ for 7 TeV and $y_{\rm q}\approx4.46$ for 13 TeV.  Figure 4(b) indeed shows the peak of large-rapidity ridge correlations shifts to higher $y_{\rm q}$ for higher collision energies, as we expect. This is because the rapidity $y$ increases with $\sqrt{s}$ at fixed $x$ and $p_\bot$, as shown in Eq. (1). Larger rapidity gluon represents source gluons which has strong correlations with radiated gluons. 

The physics of different rapidity regions represent different stages of gluon evolution in the CGC framework. This physics picture may also explain why the $v_2$ and $v_3$ are different
given by the PHENIX and STAR collaboration~\cite{PHENIX,STAR}, where their detectors cover forward and central rapidity regions, respectively.   

As Eqs. (7) and (12) demonstrate, $Y(\Delta y)$ is an integrated correlation function while $C(\boldsymbol{p}_\bot, y_{\rm p}; \boldsymbol{q}_\bot, y_{\rm q})$ is a differential correlation function. Fig.~4 demonstrates that the rebound in the integrated correlation function is modest while in the differential correlation function it is more easily observed. 

The patterns of large-rapidity ridge correlations with respect to $\sqrt{s}$ is a characteristic of the CGC mechanism. Identifying this characteristic is a possible way to test the mechanism of the CGC.

\section{summary}
In this study, within the framework of CGC, we propose an exact normalization scheme for the longitudinal rapidity correlations. In this exact scheme, the violation of boost invariance of the rapidity distribution is taken into account. The large-rapidity ridge correlation rebounds after bottoming, which is consistent with the observed data at the CMS detector. 

The rebound of large-rapidity ridge correlations is related to the quantum evolution of gluons. The physics of different rapidity regions is different within the CGC framework. Large-rapidity ridge correlations probe strong correlations between source gluons and radiated gluons. This physical picture may also understand why the $v_2$ and $v_3$ given by different rapidity regions, such as the PHENIX and STAR collaboration, are different.

The correlation rebound is further found to appear around the sum of the saturation momentum of the projectile and target. Meanwhile, the rebound moves to larger rapidities at higher colliding energies. These features are directly caused by the effect of gluon saturation in the theory of CGC. 

It is also shown that the rebound in the large-rapidity ridge correlation is more easily observed in the differential correlation function $C(\boldsymbol{p}_\bot, y_{\rm p}; \boldsymbol{q}_\bot, y_{\rm q})$. Further observation of the rebound in the differential correlation function is a direct test of the CGC mechanism.

\providecommand{\href}[2]{#2}\begingroup\raggedright\endgroup
\end{document}